\documentclass[12pt,preprint]{aastex}

\usepackage{amsmath}

\usepackage{graphicx}

\newcommand{\liemail}{litp@tsinghua.edu.cn}
\newcommand{\liuemail}{liuhao@ihep.ac.cn}
\newcommand{\degree}{^\circ}

\def \<{\langle}
\def \>{\rangle}

\shorttitle{Missing of CMB quadrupole}
\shortauthors{Liu \& Li}

\begin{document}
\title{Missing completely of the CMB quadrupole in WMAP data}
\author{Hao LIU\altaffilmark{1} and Ti-Pei LI\altaffilmark{1,2}}
\altaffiltext{1}{Key Lab. of Particle Astrophysics, Institute of High Energy Physics,
Chinese Academy of Sciences, Beijing; \liuemail}
\altaffiltext{1}{Center for Astrophysics, Tsinghua University, Beijing, China; \liemail}

\begin{abstract}
In cosmic microwave background (CMB) experiments, foreground-cleaned temperature maps are still contaminated by the residual dipole due to uncertainties of the Doppler dipole direction and microwave radiometer sidelobe. To obtain reliable CMB maps, such contamination has to be carefully removed from observed data. We have previously built a software package for map-making, residual dipole-contamination removal, and power spectrum estimation from the Wilkinson Microwave Anisotropy Probe (WMAP) raw data. This software has now been significantly improved. With the improved software, we obtain a negative result of CMB quadrupole detection with the WMAP raw data, which is $-$3.2\,$\pm$ \,3.5\, $\mu$K$^2$ from the seven-year WMAP (WMAP7) data. This result is evidently incompatible with  $\sim 1000\,\mu$K$^2$ expected from the standard cosmological model $\Lambda$CDM. The completely missing of CMB quadrupole poses a serious challenge to the standard model and sets a strong constraint on possible models of cosmology. Due to the importance of this result for understanding the origin and early evolution of our universe, the software codes we used are opened for public checking.
\end{abstract}


\section{Introduction}
The CMB quadrupole is the largest observable structure in our universe. It corresponds to the temperature fluctuation power at the lowest detectable multiple moment $l=2$, or $\sim 90\,\degree$ angular scale, and reflects the universe circumstance at very early epoch. It is a difficult task to truly evaluate the quadrupole moment from an observed microwave map. We find that, although the foreground contamination problem has been effectively treated in CMB observations, another important systematic error, the scan-induced dipole contamination accumulated on large angular scales, still exists and produces a false aligned quadrupole for all sweep missions~\citep{liu11a,liu11b}, which, unfortunately, has not been adequately realized and removed from the official results released by the CMB mission groups.

For a sky pixel with the unit direction vector ${\bf n}$, the Doppler dipole signal is
\begin{equation}\label{dip}
T_{_{d}}=\frac{T_0}{c}{\bf v}\cdot{\bf n}\,,
\end{equation}
 where $T_0=2.725$\,K is the monopole temperature, ${\bf v}$ the joint velocity of antenna relative to the CMB, $c$ the speed of light. The amplitude of Doppler dipole moment is greater than $3000\,\mu$K, much stronger than the CMB anisotropy ($\sim50\,\mu$K). The dipole signal $T_{_{d}}$ must be calculated with Eq.\,\ref{dip} for each observation and removed from the measured temperature. However, the uncertainties of ${\bf v}$ and ${\bf n}$ in Eq.\,\ref{dip} make it impossible to estimate $T_{_{d}}$ accurate enough. Firstly, the  direction error $\sim10'$ in the vector ${\bf v}$, i.e. in the CMB dipole~\citep{ben03a}, can cause the calculated dipole signal to deviate by as much as $10-20\,\mu$K, which cannot be ignored when compared to the relatively weak CMB anisotropy. Secondly, in calculating $T_{_{d}}$ with Eq.\,\ref{dip}, the effect of overall sidelobe response uncertainty of radio telescope can equivalently introduce an error in the direction vector ${\bf n}$, which is estimated to be $\sim10'-20'$ for WMAP~\citep{liu11a}. Furthermore, an undocumented 25.6\,ms timing offset error in WMAP time-ordered-data (TOD) has been revealed \citep{lxl,rou10}, which also can deviate calculated dipole signals as an antenna pointing direction error of $\sim7'$ does . Different errors in dipole calculation with Eq.\,\ref{dip} can be synthesized to an overall-equivalent direction error $\Delta{\bf n}$ with a magnitude of about $10'$. Since $\Delta{\bf n}$ includes both real and equivalent pointing error, it can not be eliminated by improving only the real antenna pointing accuracy. After a full sky scan, these $\Delta{\bf n}$ induced dipole deviations should be accumulated in the resulting CMB map, producing artificial anisotropies at the large angular scales (not only the dipole) with a pattern closely correlated to the scan pattern and an artificial aligned quadrupole~\citep{liu11b}.

The WMAP scan scheme makes its sky coverage inhomogeneous -- the number of observations being greatest at the ecliptic poles and the plane being most sparsely observed. As above expected, it is really existed in released WMAP temperature maps a significant large-scale non-Gaussian modulation feature being closely correlated to the scan pattern~\citep{spe06}, and the CMB quadrupole is oddly aligned along the scan pattern (see the first plot of Fig.~\ref{fig:template}) called the "axis of evil"~\citep{teg04, sch04, lan05}. With an assumed overall-equivalent direction error $\Delta{\bf n}$ of $7'$, we simulated the first year WMAP scan survey, calculated the dipole deviation for each observation, and found that the accumulated temperature distortion, which is generated using only spacecraft attitude information and the scan scheme without any CMB information at all, is well consistent with the CMB quadrupole component released by the WMAP team~\citep{lxl}. Therefore, without removing the accumulated dipole contamination, released CMB anisotropy at the largest scales is not reliable.

To give the exact overall-equivalent direction error $\Delta{\bf n}$ by estimating all effects causing the dipole contamination is impossible, because the sidelobe uncertainty can not be eliminated completely. Independent of the reasons that cause $\Delta{\bf n}$, we have proposed a template fitting procedure to remove the dipole contamination~\citep{liu11a}. The overall-equivalent direction error can be expressed as $\Delta{\mathbf n}=\Delta{\mathbf x}+\Delta{\mathbf y}+\Delta{\mathbf z}$ in the spacecraft coordinate system $(X, Y, Z)$,  where the $X$-axis is parallel to plane of radiators, the $Z$-axis is the anti-sun direction of the spin axis, and the $Y$-axis is perpendicular to both. For an assumed overall direction error $\Delta{\bf n}=\Delta{\mathbf x}=1'$, we can follow the WMAP observational scan history to calculate dipole deviations by $\delta T_x=(T_0/c){\bf v}\cdot \Delta{\bf n}$ for each observation and accumulate them into a template map $\Delta T_x$.  The other two template maps $\Delta T_y$ and $\Delta T_z$ can be similarly derived by assuming $\Delta{\bf n}$ of $1'$ being along the $Y$- and $Z$-axis, respectively. With template cleaning, from an observed and monopole subtracted map $T'$ the clean temperature map $T$ can be derived by
\begin{equation}\label{equ:fitting}
T=T'-(a_x \Delta T_x+a_y \Delta t_y+a_z \Delta t_z)\,,
\end{equation}
where the coefficients $a_x$, $a_y$ and $a_z$ are determined by minimizing the variance of $T$,
\begin{equation}\label{equ:min}
T^2=\min\,.
\end{equation}

The template maps $\Delta T_x$, $\Delta T_y$ and $\Delta T_z$ have to be generated from the WMAP original raw data with map-making software. To remove the accumulated dipole contamination with Eq.\,~\ref{equ:fitting} and \ref{equ:min} to produce cleaned CMB maps, we must redo the whole map-making process from WMAP raw TOD\footnote{Because the deviation templates $\Delta T_{x,y,z}$ must be calculated with the same pipeline for the temperature map-making.}. Because the WMAP map-making software has not been opened for public use, we built a self-consistent software package of map-making and power spectrum estimation.
With our previous software, we modeled and removed the pseudo-dipole signal from the released WMAP7 maps of the Q-, V- and W-bands and then from the clean maps obtained the average residual quadrupole power of $\sim 17.1\,\mu$K$^2$, only $\sim 14\%$ of what has been released by the WMAP team~\citep{liu11a}. Since then our software has been much improved, and in this work we use the improved software and obtain the amplitude of CMB quadrupole as $-$3.2\,$\pm$ \,3.5\, $\mu$K$^2$ from the WMAP7 data, which indicates that the CMB quadrupole predicted by the standard cosmology model might be completely missing.

\section{Analysis software}\label{sec:software}
Independently of the WMAP team, we developed a self-consistent software package of map-making and power spectrum estimation, which passed a variety of tests~\citep{liu09a} and used in~\cite{liu11a}.

\subsection{Improvements}
\label{sub:improvements}
The software we used in~\cite{liu11a} has many deficiencies as briefly listed below:
\begin{itemize}
  \item The antenna pointing error is a preset absolute value.
  \item It does not distinguish the difference between real and equivalent pointing errors.
  \item The temperatures inside the processing mask are not calculated.
  \item The foreground removal and power spectrum calculation have not been included as a whole package.
  \item It is not easy to use.
\end{itemize}

Now the software has been much improved:
\begin{itemize}
  \item The antenna pointing error is determined by template fitting, no preset value any more.
  \item The templates are calculated for each single year and single band separately.
  \item We consider mainly the equivalent pointing error now.
  \item The correct loss-imbalance factors are given automatically according to the TOD version (WMAP3, 5, 7).
  \item The produced maps are now full-sky.
  \item The software now starts from the TOD and ends at the final CMB cross-power spectrum.
  \item Now all intermediate steps are done automatically by typing just a short command.
\end{itemize}

It is important to note that, all improvements to the software are "fair" for quadrupole detection. This means that, in principle, they do not prefer a higher or lower quadrupole by themselves.  This will be further discussed in \S\,\ref{sec:sig}.

The current software contains the following modules, which will be discussed one by one:
\begin{enumerate}
  \item Unpack the TOD
  \item Make the CMB temperature maps
  \item Remove the foreground emission
  \item Remove the large scale deviations
  \item Compute the output CMB power spectrum
\end{enumerate}

\subsection{Modules in this version}\label{sub:modules}
The first module is to unpack the TOD. In this module, we compute the A- and B-side antenna pointing vectors $\mathbf{n_{_A}}$ and $\mathbf{n_{_B}}$, the differential Doppler signal 
\begin{equation}\label{equ:dipole}
d=\frac{T_0}{c}\mathbf{v}\cdot(\mathbf{n_{_A}}-\mathbf{n_{_B}})
\end{equation}
with $T_0$ being the CMB monopole, $c$ the light speed, $\mathbf{v}$ the velocity of the spacecraft relative to the CMB rest frame, and the differential dipole deviations 
\begin{equation}\label{equ:time-order deviation}
d^*_{x,y,z}=\frac{T_0}{c}\mathbf{v}\cdot \mathbf{\Delta n}_{_{x,y,z}}
\end{equation}
with $\Delta\mathbf{n}_{_x}$, $\Delta\mathbf{n}_{_y}$, $\Delta\mathbf{n}_{_z}$ being the equivalent pointing errors along the $X$, $Y$, $Z$ axis of the spacecraft coordinate (see Eq.\,4 of \cite{liu11a}). Since they will eventually be determined by fitting, the absolute amplitudes of $\Delta\mathbf{n}_{_x}$, $\Delta\mathbf{n}_{_y}$, $\Delta\mathbf{n}_{_z}$ are not important here. Unlike the previous version, in this module we do not apply any data selection. All data will be passed to the next module.

The second module is to make CMB temperature maps. Here we exclude some of the data by TOD flags and planet position, and make the output CMB temperature maps according to Eq.\,19 of~\cite{hin03} with 80 rounds of iterations. The default is to make single year maps, because the CMB cross power spectrum should be derived from single year maps. The main difference to the previous version is: The data selection is done here, so we do not have to re-unpack the TOD for a different data selection criterion. This saves a lot of time because the unpacking is the most time-consuming part of all. In this step, we also make three temperature deviation templates $\Delta T_x$, $\Delta T_y$ and $\Delta T_z$ from $d^*_x$, $d^*_y$ and $d^*_z$, with exactly the same condition to the corresponding single year CMB temperature map. The masked region is first ignored, and then, after other regions have been generated, they are used to re-calculate the masked region to give a full-sky output. Due to the planet exclusion, some places might still be empty; however, they are no more than $0.1\%$ of the full sky, so we just fill them with zero to obtain a full sky output, and set the number of observations of those pixels to zero so that they can be easily identified.

After making maps, we then smooth them and remove the foreground according to \cite{ben03b}. In previous work, we simply use the WMAP coefficients for foreground subtraction; however, this time the coefficients are self-consistently determined by data, which is apparently more reliable.

The next step is to remove $\Delta T_x$, $\Delta T_y$ and $\Delta T_z$ from the corresponding single year  map $T'$ by template fitting with Eqs.\,\ref{equ:fitting} and \ref{equ:min} to get the foreground cleaned and dipole deviation fixed CMB temperature map $T$. In previous version, considering the correlation between $\Delta T_x$, $\Delta T_y$, $\Delta T_z$, we use only $\Delta T_x$ in fitting. This time, we first remove the components that are proportional to $\Delta T_x$ from $\Delta T_y$ and $\Delta T_z$, and then remove the component that is proportional to the remaining $\Delta T_y$ from the remaining $\Delta T_z$ to get three linearly independent templates, which are removed from $T'$ by fitting.

In the last step, we calculate all CMB power spectra from the cross of all single year CMB temperature maps using PolSpice~\citep{polspice2,polspice1}, and then bin the average of all cross power spectra according to the WMAP binning scheme. Both binned and unbinned power spectra are provided as the final outputs.

\subsection{Power spectra}
Using our software we produce CMB temperature anisotropy maps $T(\mathbf n)$ from WMAP TOD data, and decompose $T(\mathbf n)$ in spherical harmonics
\begin{equation}\label{equ:decompose}
T(\mathbf n)=\sum_{l>0}\sum_{m=-l}^{l}a_{_lm}Y_{lm}(\mathbf n)
\end{equation}
where $\mathbf n$ is a unit direction vector. For different multiple moment $l$ we calculate the angular power $l(l+1)C_l/2\pi$ where
\begin{equation}\label{equ:cl}
C_l=\frac{1}{2l+1}\sum_{m=-l}^{l}|a_{_{lm}}|^2\,.
\end{equation}

\begin{figure}[t]
\begin{center}
\includegraphics[height=4.cm, angle=90]{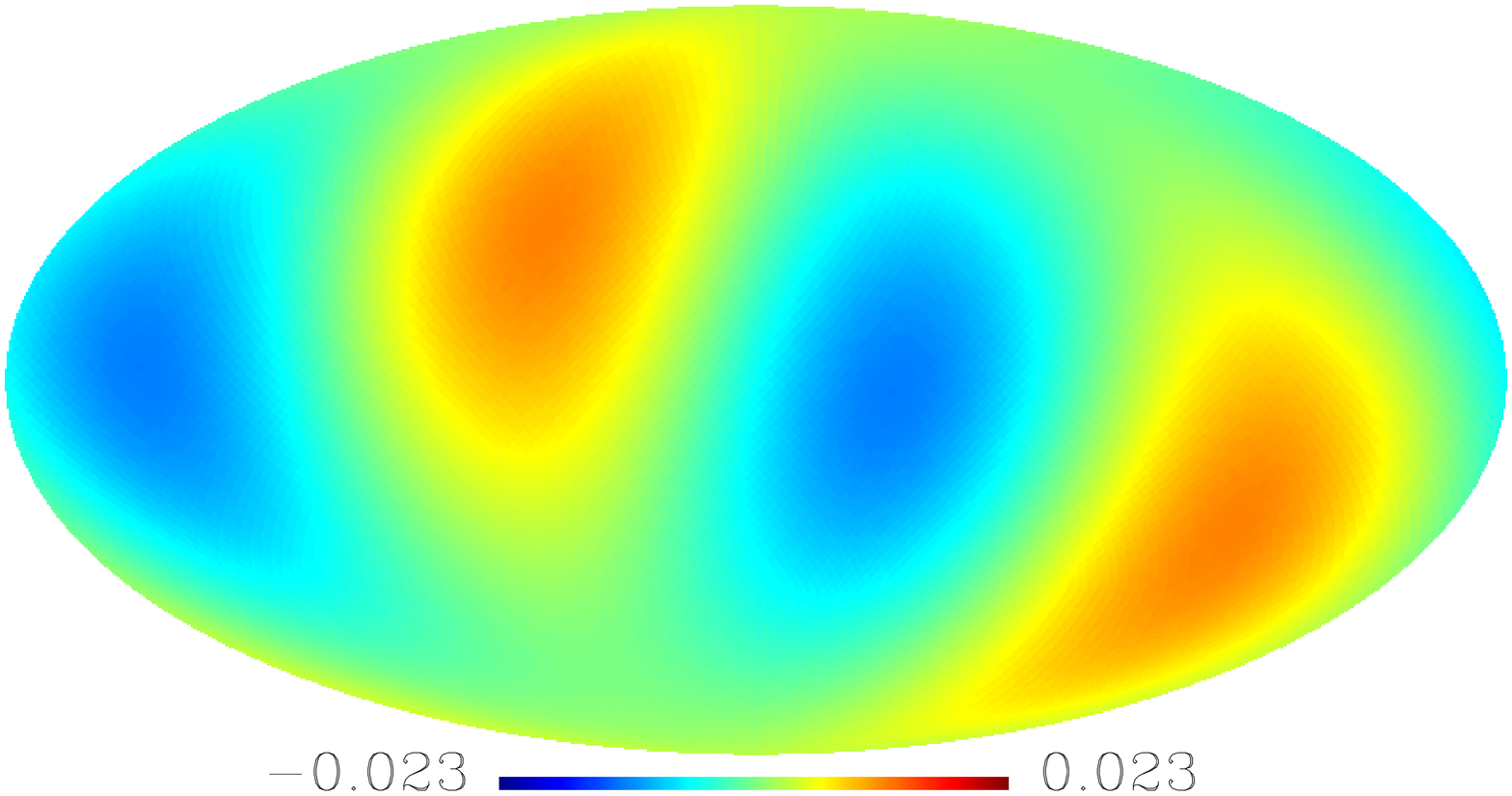}
\includegraphics[height=4.cm, angle=90]{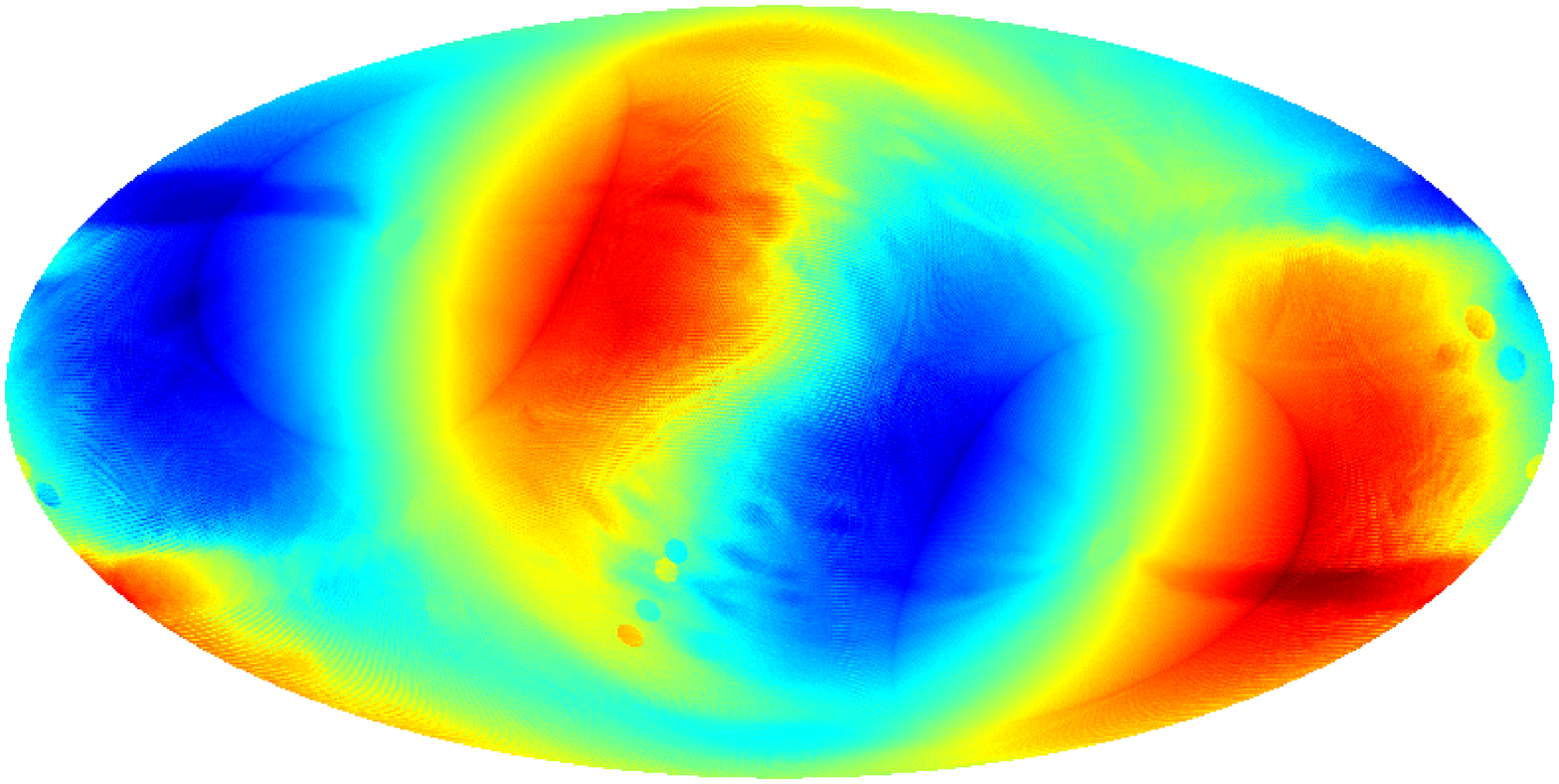}
\includegraphics[height=4.cm, angle=90]{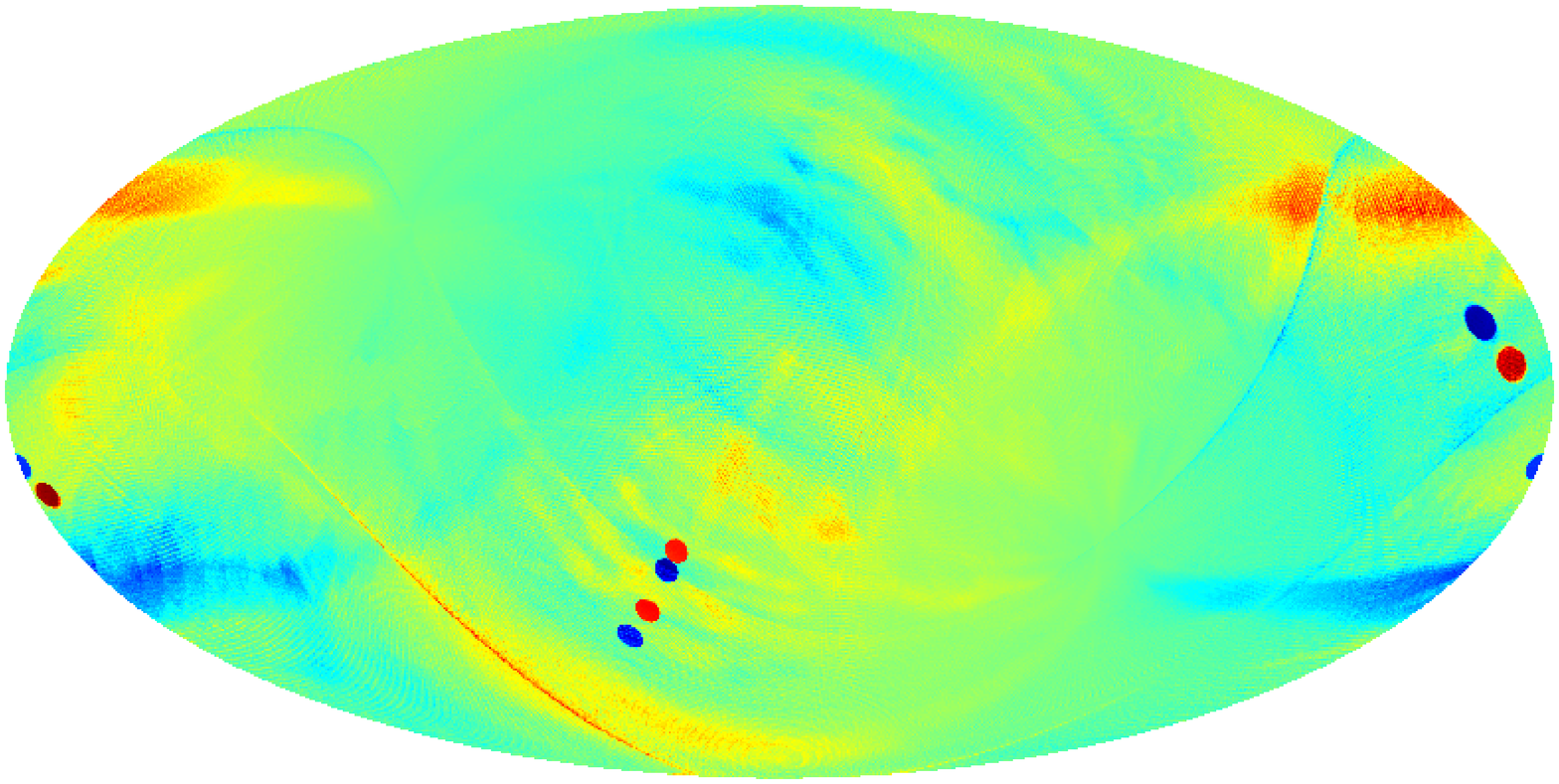}
\includegraphics[height=4.cm, angle=90]{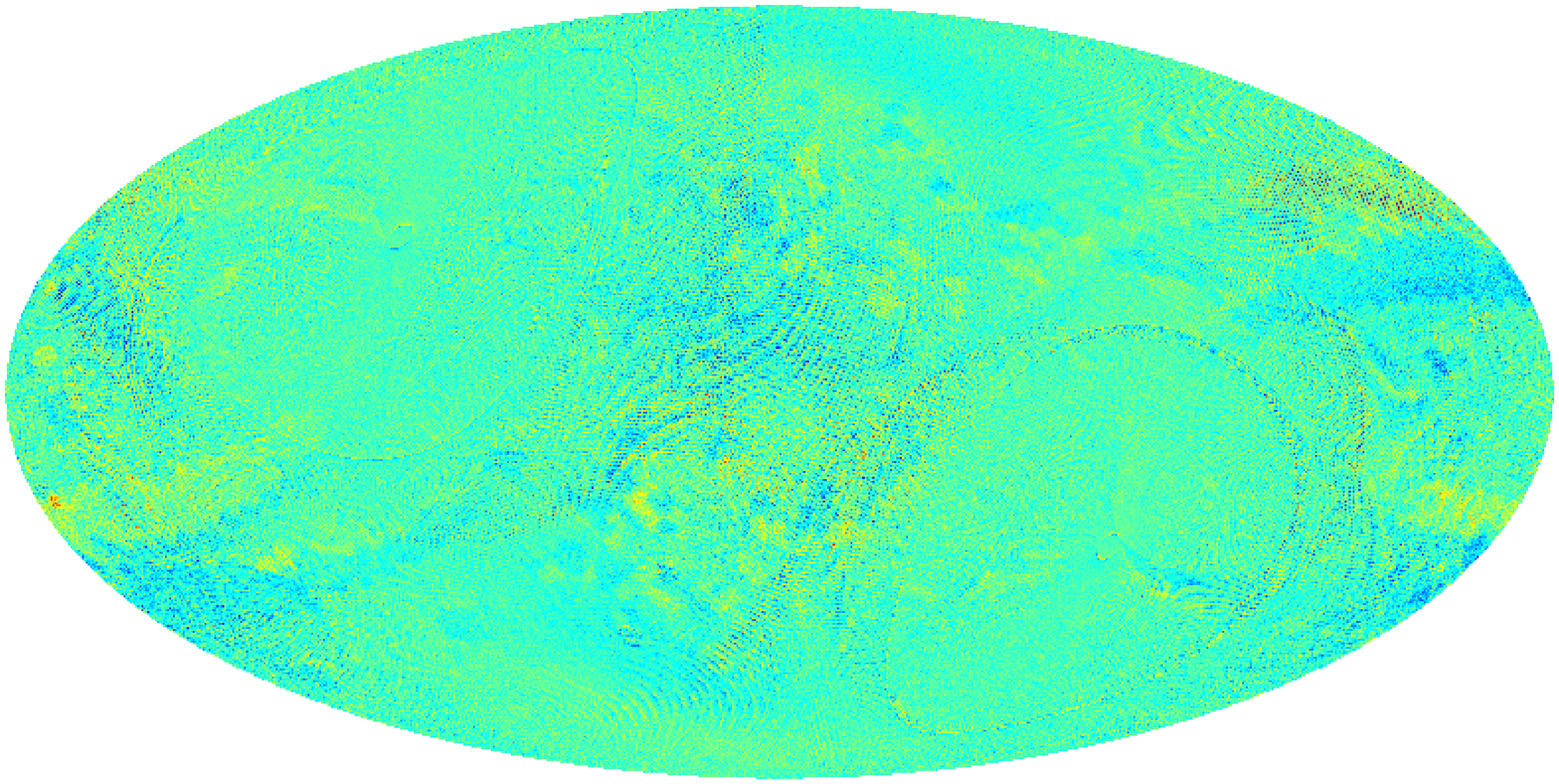}
\end{center}
\vspace{-7mm}
\caption{From left to right, the first plot is the released CMB quadrupole component derived from WMAP5 V and W bands. The other three plots are examples of the X-, Y-, and Z- templates (WMAP7, year-1, W1-band) used in our new code to remove deviations due to the overall-equivalent direction error. The new templates are significantly improved: They are whole-sky maps, and each single-year, single-band map has its own X- Y- Z- templates. All plots are in Galactic coordinates.
}\label{fig:template}
\end{figure}

\section{Analysis result}\label{sec:result}
For each single year WMAP observation and each band, we calculate the accumulated dipole contamination maps $\Delta T_x$, $\Delta T_y$ and $\Delta T_z$ as removal templates with an assumed equivalent pointing error $\Delta\mathbf{n}_{_x}=\Delta\mathbf{n}_{_y}=\Delta\mathbf{n}_{_z}=1'$. As an example, the later three plots of Fig.~\ref{fig:template} show the obtained template maps $\Delta T_x$, $\Delta T_y$ and $\Delta T_z$ from the WMAP7 year-1 W1-band data.

Now we can remove the accumulated dipole deviation with our map-making software. We produce single-year CMB temperature maps from the WMAP3, WMAP5 and WMAP7 data separately, and clean for the foreground as done by WMAP, and remove the dipole contamination by template fitting by Eqs.\,\ref{equ:fitting} and \ref{equ:min}. After that, cross-power spectra are calculated from each pair of the clean maps, and the average of all cross-power spectra is used as the final CMB power spectrum. 

\begin{figure}[t]
\includegraphics[width=0.35\textwidth,angle=270]{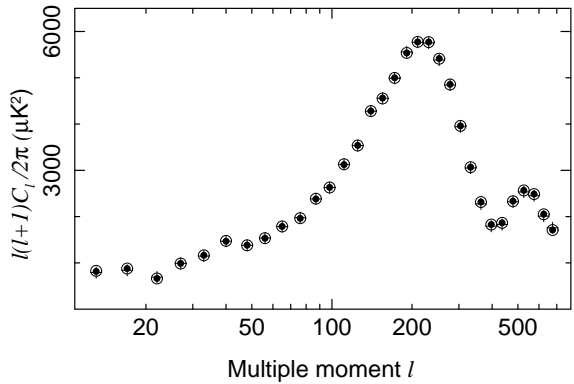}
\caption{High $l$ CMB angular power spectra. {\sl Cross}: the WMAP7 official release. {\sl Circle}:
the spectrum derived from WMAP7 data using our software without removing dipole deviation. {\sl Filled circle}: the spectrum derived from WMAP7 data by our software after removing dipole deviation.}
\label{fig:high}
\end{figure}

From Fig.~\ref{fig:high}, we notice that our resulting spectrum remains highly consistent with the WMAP release at high-$l$, either before or after the dipole contamination removal. This means that our software is fully consistent to WMAP and the deviation removal which affects mainly the low-$l$.

\begin{figure}[t]
\includegraphics[width=0.4\textwidth,angle=270]{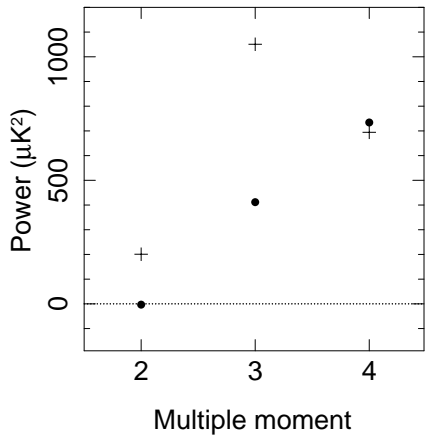}
\caption{Low $l$ CMB power spectra. \emph{Filled circle}: our result from WMAP7 data after the dipole contamination removal.  \emph{Cross}: the WMAP7 official release. The unit is $\rm{\mu K^2}$. All the error bars in this figure are small than the marker symbols.}
\label{fig:low}
\end{figure}

Unlike the high-$l$, the quadrupole and octopole results are significantly different to the WMAP release.
The amplitudes of CMB quadrupole from WMAP3, WMAP5 and WMAP7 data drop from the WMAP release of greater than 200 $\mu$K$^2$ to between $-4$ and $-0.1$~$\mu$K$^2$, as shown in Fig.~\ref{fig:low} and Table~\ref{tab:l23}. The amplitude error mainly consists of two components: map-to-map variance (due to noise, residual foreground, etc) and incomplete sky coverage (due to the foreground mask). The map-to-map variance can be directly derived from the cross-power spectra\footnote{Each independent single-year, single-band temperature map will be used repeatedly in calculating the cross power spectra; however, this does not significantly affect the variance calculation, because each single-map has the same contribution in the final cross power spectra.}, and the incomplete sky coverage effect can be estimated by simulations. In this way, the standard error of the cleaned CMB quadrupole of WMAP7 is estimated to be 3.5\,$\mu$K$^2$, very close to the standard error give by the WMAP team here ($\sim5\rm{\mu K}^2$). Therefore, the result of CMB quadrupole detection from the seven year WMAP observation can be reported as $Q=-3.2\pm3.5\,\mu$K$^2$.

\begin{table}
\caption{CMB quadrupole and octopole  ($\rm{\mu K^2}$)}
\begin{center}
\begin{tabular}{c|c c|c c}
  \hline
   Data   &\multicolumn{2}{c}{Quadrupole} &\multicolumn{2}{c} {Octopole} \\
\cline{2-5}
       & WMAP release & This work & WMAP release     & This work    \\
\hline
  WMAP3  &     211   & -0.1  & 1041 & 430 \\

  WMAP5   &    213   & -4.0  &1038 & 415 \\

  WMAP7    &   201 & -3.2 &1050 & 411 \\
  \hline
\end{tabular}
\end{center}
\label{tab:l23}
\end{table}

\section{Statistical significance}\label{sec:sig}
\subsection{The degree of freedom problem}\label{sub:dof}
In stead of only one template $\Delta T_x$ used in \cite{liu11a,liu11b} to remove pseudo-dipole signal, now we use all three templates $\Delta T_x$, $\Delta T_y$ and $\Delta T_z$. Since the quadrupole ($l=2$) has only five degrees of freedom (see Eq.~\ref{equ:decompose} and \ref{equ:cl} where $m$ from -2 to 2), theoretically speaking, if we have five linearly independent quadrupole templates, it will be possible to remove the quadrupole completely from any map by using Eq.~\ref{equ:fitting}. Although we have only three templates: $\Delta T_x$, $\Delta T_y$ and $\Delta T_z$, one may still worry about the degree of freedom problem.

\begin{figure}[t]
\includegraphics[width=0.4\textwidth]{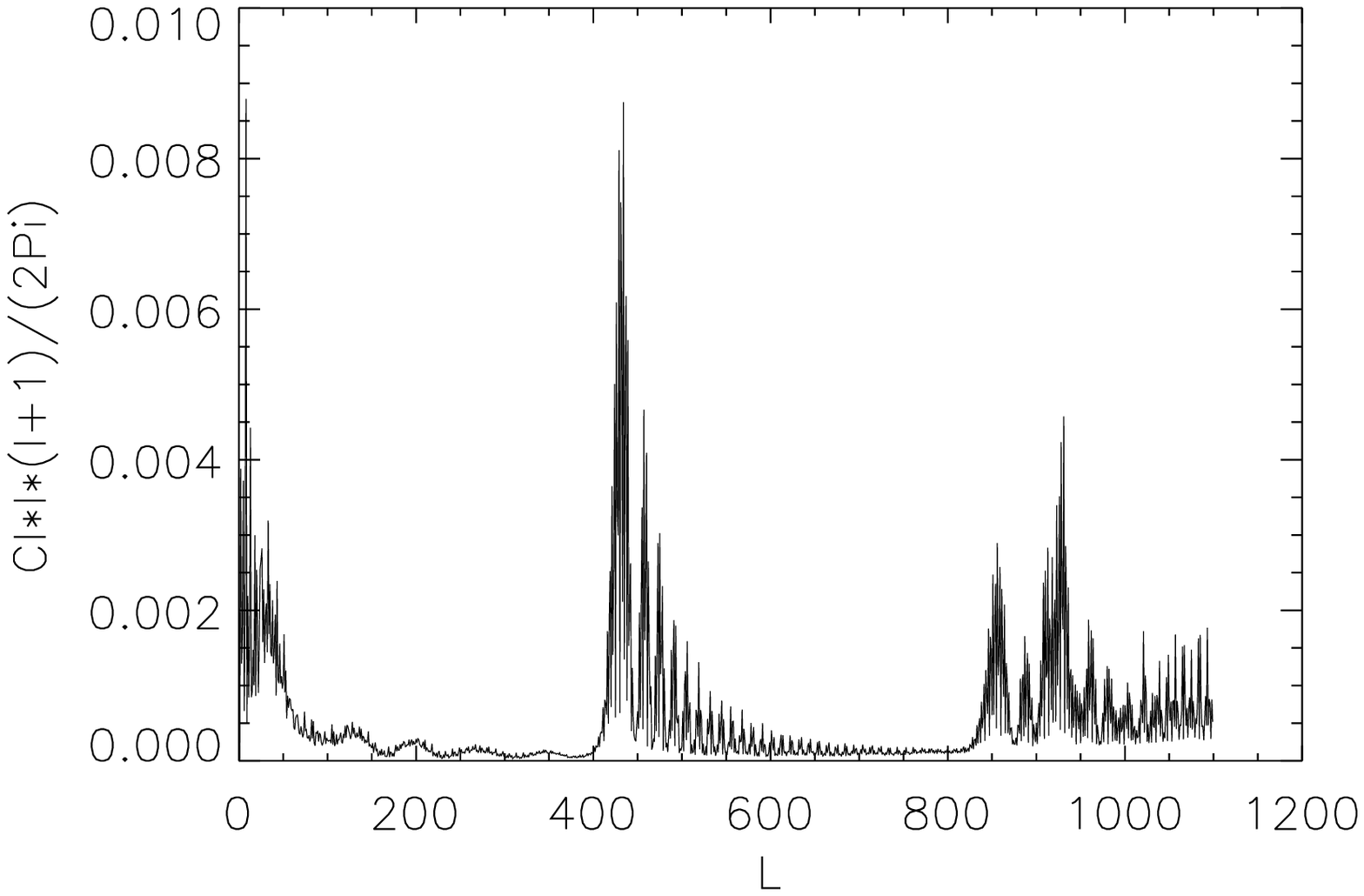}
\includegraphics[width=0.4\textwidth]{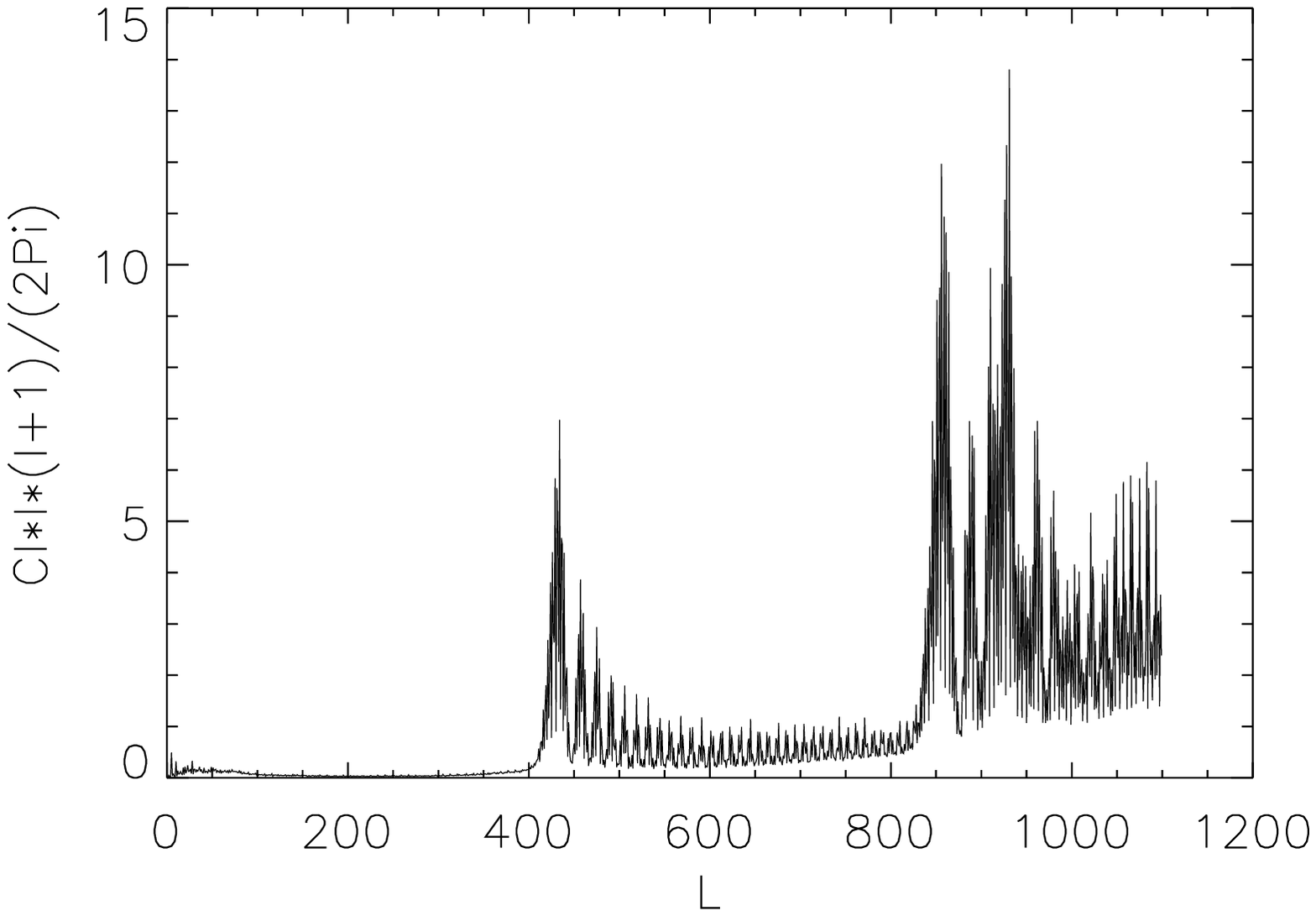}
\vspace{-4mm}
\caption{The power spectra of  $\Delta T_y$ (left) and $\Delta T_z$ (right) templates of year-1, W1. The dipoles and monopoles have been removed in advance, and the correlated components between $\Delta T_x$, $\Delta T_y$ and $\Delta T_z$ have also been canceled as illustrated in \S\,\ref{sub:modules}. The maximum power for $\Delta T_y$ occurs at $l=8$, and the maximum power for $\Delta T_z$ occurs at $l=931$. }
\label{fig:dy and dz}
\end{figure}

However, it must be noticed that, the precondition for degree of freedom problem is that the templates must be quadrupole templates, or at least dominated by the quadrupole. This is true for the $\Delta T_x$ template, but false for $\Delta T_y$ and $\Delta T_z$ templates. As shown in Fig.~\ref{fig:dy and dz}, both  $\Delta T_y$ and $\Delta T_z$ are dominated by non-quadrupole components. Therefore, when we add these two templates into fitting, the degree of freedom will not change. In other words, by adding these two templates into fitting, the probability distribution of the resulting quadrupole power will not change significantly.

\begin{figure}[t]
\includegraphics[width=0.4\textwidth]{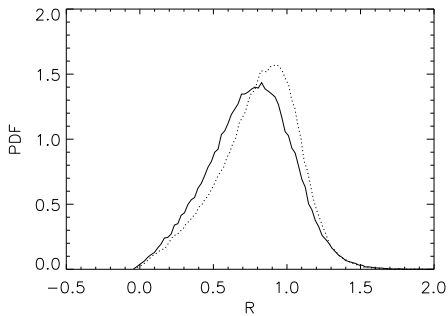}
\vspace{-5mm}
\caption{The probability density functions of $R = C_2/C'_2$, where $C'_2$ is derived before template fitting, and $C_2$ after fitting. \emph{Solid line:} all $x$-, $y$- and $z$-templates are used in fitting. \emph{Dotted line:} only the $x$-template is used in fitting.}
\label{fig:rate after fit}
\end{figure}

This can be tested by simulation: We generate $3\times10^5$  simulated CMB temperature maps according to the best-fit WMAP CMB power spectrum, with the CMB quadrupole $C_l l(l+1)/2\pi$ ($l=2$) set to $100$ $\rm{\mu K^2}$, and apply the deviation template fitting with the same process applied to the real data, then compute the ratio $R = C_2/C'_2$, where $C'_2$ is obtained before fitting and $C_2$  after fitting. The distribution of $R$ for only one template $\Delta T_x$ used is shown by the dotted line in Fig.~\ref{fig:rate after fit}, and that for three templates by the solid line.  It is apparent that, by adding the $y$- and $z$-templates into fitting, the probability density function does not change significantly. The average of $R$ is 0.761 for three templates condition, and 0.813 for $x$-only condition, very close to each other. We can also see that for both case, many simulated map can give $R>1$, which means the fitting even has a good chance to increase the quadrupole.

We should also notice that, with either methods ($xyz$ or $x$-only), the average of $R$ will be systematically lower than 1. This is inevitable due to chanced correlation between the templates and true CMB - when we try to remove the foreground with emission templates, the same thing also happens. We never deny such a deficiency. However, as we can see from simulation, the average of $R$ is close to 1. This tells us that, before finding a better method to calculate the exact accumulated dipolar error, it is still necessary to clean the contamination using Eqs.~\ref{equ:fitting} and \ref{equ:min}, and keep aware of the possible deficiency, like what we have done in cleaning the foreground.

\subsection{What does the negative quadrupole mean?}\label{sub:negative quadrupole}
Another problem is the CMB quadrupole value after removing dipole contamination being negative. However, from Eq.\,\ref{equ:cl} it must be greater than or equal to zero. Incomplete sky coverage due to the galactic mask causes uncertainty to the final power spectrum. Therefore, if the true CMB quadrupole is zero or very low, we have equal chance to get positive or negative estimations, otherwise the estimation will be biased. In other words, with the existence of the Galactic mask and zero/nearly zero quadrupole, negative quadrupole estimation is just a natural result of statistical uncertainty.

To correctly understand the statistical meaning of a negative $C_2$, the best way is to use it to find the "most likely" true quadrupole power ($C_2^{true}$). This means: Given the CMB quadrupole value estimated from the experimental data set ($C_2$, can be negative), what's the most likely true CMB quadrupole ($C_2^{true}$, must be positive or zero)? This can be done by using the Bayes's theorem:
\begin{equation}\label{equ:Bayes for most likely c2}
P(C_2^{true}|C_2)=\frac{P(C_2|C_2^{true})P(C_2^{true})}{P(C_2)}
\end{equation}
Where $P(C_2^{true}|C_2)$ is what we want: the conditional probability of the true quadrupole power given $C_2$, and $P(C_2|C_2^{true})$ is the conditional probability of getting $C_2$ given a true CMB quadrupole, which can be obtained by simulation.

In general case, it is quite difficult to calculate Eq.~\ref{equ:Bayes for most likely c2}; however, here the problem can be greatly simplified: For a given data set, $C_2$ is a constant, so do $P(C_2)$. $P(C_2|C_2^{true})$ is a monotone decreasing function when the constant $C_2<0$. As for $P(C_2^{true})$, in most cases, when there is no a priory limitation upon $C_2^{true}$, it is a convention to assume $P(C_2^{true})=1$. Now we can say that, when $C_2<0$, Eq.~\ref{equ:Bayes for most likely c2} will be maximized at $C_2^{true}=0$. Stronger negative value for $C_2$ will only make $P(C_2^{true}|C_2)$ decrease more quickly along the positive $X$ axis. Therefore, although we can not give the exact value of $P(C_2^{true}|C_2)$, we can conclude that its most likely value of $C_2^{true}$ is exactly zero.

\subsection{On the MCL method}
The maximum likelihood-$C_l$ (MCL) method is preferred in estimating both the low-$l$ CMB power spectrum and the model-data agreements in the WMAP results~\citep{Ef04,hin07,Dunkley09,Ef10,Larson11}. The measured CMB quadrupole of $\sim 120$\,$\mu$K$^2$ from the WMAP7 data has been significantly increased up to $\sim 200$\,$\mu$K$^2$ after using MCL by the WMAP team~\citep{ben11}. The core of all forms of MCL is the covariance matrix $C=<xx^T>$ and its inverse matrix $C^{-1}$. $C$ can also be written in spherical harmonic space as $C_{ll'}=\delta(l,l')C_l$ (ignoring the noise, $C_l$ is the "true" CMB power spectrum). In general cases, this is fine; however, now we have seen the possibility that the true CMB quadrupole might be zero, which can make $C^{-1}$ non-existent. Another thing also need to be noticed: The MCL method can never give a negative result for the low-$l$ power spectrum, because negative result has no likelihood meaning. However, as we discussed in \S~\ref{sub:negative quadrupole}, when there is a mask and the "true" quadrupole is suspected to be zero, negative estimators is necessary to ensure that the expectation of estimated power spectrum is unbiased (all positive estimators will only give a $>0$ average as expectation). Therefore the MCL method needs to be further studied.

\subsection{Significance}
To estimate the significance of the negative result of CMB quadrupole detection, we generate simulated $\Lambda$CDM expected CMB sky maps with the synfast routine in the HEALPix software package~\citep{gor05}. For each simulated map, we calculate the magnitude of quadrupole $Q'$ without dipole contamination cleaning and $Q$ after the template cleaning with Eq.\,\ref{equ:fitting} and Eq.\,\ref{equ:min}. From $3\times10^5$ simulated maps with $Q'=100\,\rm{\mu K^2}$, we find that only $4.1\times10^{-4}$ of them have $Q < 0.1$ $\rm{\mu K^2}$, and only $1.7\times10^{-5}$ of them have $Q < -3.2$ $\rm{\mu K^2}$. Therefore, the chance that the negative result of CMB quadrupole detection with WMAP comes from the real CMB quadrupole being occasionally identical with the accumulated dipole deviation is very low.

\section{Discussion}\label{sec:discuss}
According to $\Lambda$CDM, the lambda dark matter model with an initial inflation, the primordial density fluctuations have a nearly scale-invariant spectrum. The CMB quadrupole represents spatial temperature fluctuation on large angular scales of $\sim90\degree$. From the first-year spectrum of CMB anisotropies the WMAP team derived an CMB quadrupole of $\sim 200$\,$\rm{\mu K^2}$, which is unexpectedly low than $\sim 1000$\,$\rm{\mu K^2}$ predicted by $\Lambda$CDM~\citep{spe03}. Since the first release of WMAP results, quite a few models have been proposed to explain the strange loss of power by attributing new physics at work in the early universe. As a noticeable example, a small universe with a specific rigid topology was suggested to account the low quadrupole amplitude~\citep{lum03}. Whereas now after removing the dipole contamination the quadrupole is further missing completely in WMAP data, which  is hard to be explained along by modifying the geometry or inflationary model of early universe and poses a serious challenge to current cosmological models.

The measurement of very weak anisotropies of CMB temperature is an extremely complicated and hard task. As the first mission with unprecedented accuracy able to make the study of cosmology with high precision, it is needed time and common effort by cosmology society to reveal unrealized systematic and statistical errors in the experiment step by step. Thanks to the WMAP team for making WMAP data publicly available, from reanalyzing released WMAP raw data we realize the previously ignored problem of artificial CMB anisotropies caused by coupling of dipole deviation and observational scan and propose the method to remove it from observed data of CMB missions~\citep{liu11a,liu11b}.  Due to the importance of the missing of CMB quadrupole for understanding the origin and early evolution of our universe, the software codes we used are opened for public checking and using\footnote{The software we used in this work can be found at http://dpc.aire.org.cn/wmap/09072731/release\_v2}.

Besides the dipole contamination, we have discovered other kinds of systematic error in released CMB maps, e.g., scan-rings of hot sources being significantly cooled by $\sim 10\,\mu$K, and an up to $\sim 20\,\mu$K systematic deviation of average temperature in Galactic latitude distributions~\citep{liu09a}, temperature distortions up to $\sim 20\,\mu$K caused by the transmission imbalance of radiometers~\citep{lietal09}. These remarkable observational effects should distort powers at larger multiple moments and have to be removed from the CMB maps as well. For this purpose, besides the WMAP raw data, the analysis software is also needed for understanding the reasons of these discovered systematics and finding approaches to remove them. It will be helpful if the WMAP data processing package can be also publicly opened.

\acknowledgments  This work is supported by National Basic Research Program of China (Grand No. 2009CB824800), the National Natural Science Foundation of China (Grant No. 11033003), National Natural Science Foundation for Young Scientists of China (Grant No. 11203024), and the Youth Innovation Promotion Association, CAS. The data analysis made use of the WMAP data archive (http://lambda.gsfc.nasa.gov/product/map/).

\end{document}